# Electric field- induced Pinch-off of a migrating compound droplet in confined micro channel

Somnath Santra[1], Sayan Das[1] and Suman Chakraborty[1,*]

[1]*Department of Mechanical Engineering, Indian Institute of Technology Kharagpur, Kharagpur, West Bengal - 721302, India*

The present study looks into the pinch-off dynamics of a compound droplet, which is suspended in another fluid in a parallel plate microchannel. The droplet is subjected to a transverse electric field in the presence of an imposed pressure driven flow. For the present study, a leaky dielectric model have been taken into consideration. When a concentric compound droplet migrates in a pressure driven flow, the inner droplet shifts from the concentric position and forms a eccentric configuration that finally leads to the rupture of the outer shell. The present investigation have uncovered that the temporal evolution of droplet eccentricity as well as the kinetics of the thinning of the outer droplet are markedly influenced by the strength of the electric field as well as the electric properties of the system. The present study also shows that the conversion of different modes of droplet pinch-off mode , such as the equatorial cap breaking-off or the hole-puncturing mode can be attained by altering the electric field strength and its electrical properties. Finally the present study depicts that these factors also alter the pinch-off time as well as its location on the outer interface. Therefore the outcomes of the present study offers an effective means of modulating the morphology of compound droplets in a confined channel by applying an electric field.

* E-mail address for correspondence: suman@mech.iitkgp.ernet.in

## I. INTRODUCTION

The electric field modulated motion and deformation of droplet have been an area of long-standing interest due to its applications in wide range of technologically driven processes [1–6]. While the single droplets have frequent application in different flow processes, compound droplets also finds its potential application in growing number of emerging applications. The compound droplet (also termed as a double emulsion) has a complex multiphase structure, where the inner droplet is encapsulated in another immiscible fluid. The special morphology of a compound droplet makes it an ideal candidate for applications in material processing [7,8], pharmaceutical [9,10] and biomedical industry [11,12]. Now a days, a double emulsion has been used in distortion and recovery of a white cell migrating in plain poiseuille flow. In this application, the compound droplet replicates the dynamics of a leukocytes, where the core (inner droplet) and the shell (outer droplet) represent the cell nucleus and cytoplasm respectively [13–15]. In some of the aforementioned examples, the break up modes of the compound droplets determines their utility where as in some other cases, the stability of the compound droplets guarantees their applicability. In such types of applications, electric field provides an effective means of manipulating the dynamics of compound droplets.

In the absence of electric field, when a concentric compound droplet is subjected to background pressure driven flow, the outer droplet deforms in the direction of flow and the inner droplet deforms perpendicular to the flow direction [16]. The degree of deformation is denoted by the relative strength of viscous stress over capillary stress characterized by capillary number *Ca*. Along with deformation, the inner droplet also shifts from its concentric configuration, therefore an eccentric configuration is developed. Due to the opposite deformation as well as the faster motion of the inner droplet, a thin region is created, where the gap between the interfaces is minimum. Eventually the gap becomes negligible and the outer droplet ruptures to release the inner droplet [16]. In a related study, Borthakur et al. (2018) have shown that the thinning rate (rate of decrease of the gap between the inner and outer interface) follows a power law criteria, where, at the initial stage of droplet motion, the thinning occurs very rapidly. On the other hand, before the occurrence of pinch off, the thinning occurs very slowly. The patterns of deformation of the compound droplet in pressure driven flow are far from being obvious when electric field is present [17–24]. Briefly, in presence of electric field, electric stress are generated at the droplet interfaces that causes the deformation of it. The electrical parameters that play prime role in governing the deformation of the interfaces are conductivity ratio, $R_{12}=\sigma_1/\sigma_2$, $R_{23}=\sigma_2/\sigma_3$ and permittivity ratios, $S_{12}=\varepsilon_1/\varepsilon_2$, $S_{23}=\varepsilon_2/\varepsilon_3$, where $\sigma$ and $\varepsilon$ are the conductivity and permittivity of the system and subscript *1*, *2* denote the inner and outer droplet phase respectively. Depending on the values of conductivity and permittivity ratios, the outer-inner droplet shows four possible types of deformation; i.e prolate-prolate, prolate-oblate, oblate-oblate and oblate prolate that can again alters the pinch-off phenomenon of the system.

Motivated from the above observation, we are interested to study the pinch-off phenomenon of compound droplet migrating in a pressure driven flow in confined microchannel under transverse electric field. To the best of the our knowledge, this investigation is yet to be done. In the present analysis, our objectives are to study the effect of electric field and electrical parameters on the (i) temporal evolution of droplet eccentricity, (ii) different modes of droplet pinch off, (iii) alteration of pinch-off time and (v) the location of the pinch off. The reported observation finds its utility in food, cosmetics, pharmacology and separation science.

## II. PROBLEM FORMULATION

### A. System description

In the present analysis, we have considered a system as shown in Fig. 1, where a concentric compound droplet is migrating in a confined pressure driven flow under transverse electric field. In the present system, all the phases are considered to be Newtonian, incompressible and leaky dielectric in nature. The undeformed radius of inner and outer droplets are $a_1$ and $a_2$ respectively. Fluid properties like viscosity, electrical permittivity and electrical conductivity are denoted by $\mu$, $\varepsilon$, and $\sigma$ respectively. The surface tension is denoted by $\gamma$. Subscript 1, 2 and 3 denote the inner phase, outer phase and the suspending phase respectively. On the other hand, 12 and 23 denote the inner and outer interface respectively.

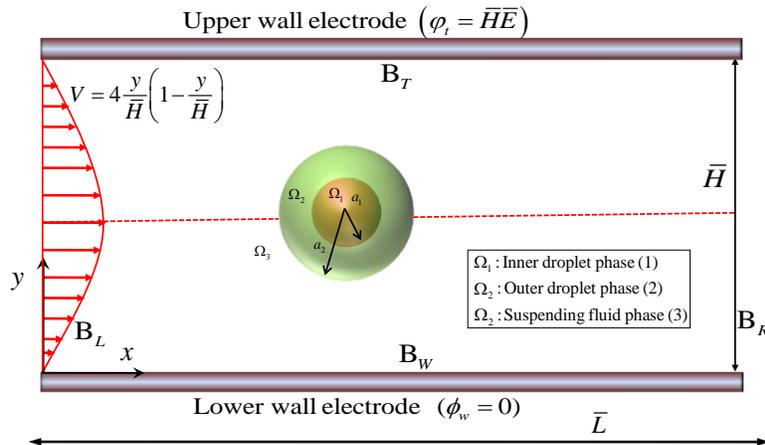

FIGURE 1. Schematic of the computational domain, where a concentric compound droplet is subjected combined presence of pressure driven flow and transverse electric field. The electric potential at the upper and lower electrode are $\bar{H}\bar{E}$ and 0 are respectively.

The distance between the upper and lower electrode wall is $\bar{H}$ and the strength of electric field, directed perpendicular to the direction of flow is denoted by $\bar{E}$. Therefore the electric potential at the top and bottom electrode are $\bar{H}\bar{E}$ and 0 respectively.

### B. Governing equation with Phase field method

*1. Mathematical challenges and solution methodology*

The present mathematical model is non-linear and coupled that does not allow us to get a exact analytical solution of the present problem for random values of the electrical parameters. Further complexity arises from the consideration of confined domain. Towards capturing the essential physics of EHD modulated dynamics of compound droplet in confined micro channel under back ground pressure driven flow, we have performed a numerical simulation. In the following section, we have discussed about the details of the numerical simulation. For non-dimesionalization of the governing equation, we have used the following characteristic scale: length $\sim \bar{H}$ (channel height), velocity$\sim u_c$ (centerline velocity), time $t_c \sim \bar{H}/u_c$, viscous stress $\sim \mu_3 u_c/\bar{H}$, electric field strength $\sim \bar{E}$ and electric stress $\sim \varepsilon_3 \bar{E}^2$. After applying these non-dimensional scheme, we have obtained some important non-dimensional number those are: capillary number, $Ca = \mu_3 u_c/\gamma_{23}$ (denotes the relative strength of viscous stress over capillary stress), electric capillary number, $Ca_E = \varepsilon_3 \bar{E}^2 \bar{H}/\gamma_{23}$ (which represents the ratio of electric stress over the capillary stress), Reynolds number, $Re = \rho u_c \bar{H}/\mu_3$ (which symbolizes the relative strength of viscous stress over stress due to inertia). Furthermore, we have also identified some non-dimensional properties ratios such as permittivity ratio, $S_{ij}=\varepsilon_i/\varepsilon_j$; conductivity ratio, $R_{ij}=\sigma_i/\sigma_j$ and viscosity ratio, $\lambda_{ij}=\mu_i/\mu_j$. Along with this, $Wc=2a_2/\bar{H}$ (ratio of outer droplet diameter and total channel height) denotes the channel confinement ratio and $K= a_1/a_2$ represents radius ratio.

*2. Numerical solutions - phase field formalism*

we have used the diffuse-interface based phase field model for the numerical simulation of the present problem [25–27]. In phase field method, the diffuse interface takes the place of the sharp interface. Therefore there is no necessity of tracking the interface. For the convenience of the numerical simulation, we have considered that the inner and the suspending phase are same. In phase field model, two different immiscible fluids are characterized by a phase field parameter $\phi$. In case of inner fluid (fluid 1) and the suspending fluid (fluid 2), we have taken $\phi(\bar{\mathbf{x}},\bar{t}) = -1$. On the other hand $\phi(\bar{\mathbf{x}},\bar{t}) = +1$ is taken for the fluid 2 (outer droplet). At the interface, the $\phi(\bar{\mathbf{x}},\bar{t})$ varies from -1 to 1 rapidly. The Cahn-Hilliard equation has been used to govern the transient evolution of $\phi(\bar{\mathbf{x}},\bar{t})$ and the equation is read as [25, 27–29]

$$\frac{\partial \phi}{\partial \bar{t}} + \bar{\mathbf{u}} \cdot \bar{\nabla} \phi = \bar{\nabla} \cdot \left( \bar{M}_\phi \bar{\nabla} \bar{G} \right), \qquad (1)$$

where $\bar{M}_\phi$ and $\bar{G} = \gamma(\phi^3 - \phi)/\bar{\zeta} - \gamma \bar{\zeta} \bar{\nabla}^2 \phi$ symbolize the interface mobility factor and chemical potential factor respectively. The interface thickness is controlled by the parameter $\bar{\zeta}$. In non-dimensional form, the Cahn-Hilliard equation is expressed as [27,28]

$$\frac{\partial \phi}{\partial \bar{t}} + \mathbf{u} \cdot \nabla \phi = \frac{1}{Pe} \nabla^2 G. \tag{2}$$

The dimensionless entities in equation (2) are $G = (\phi^3 - \phi)/Cn - Cn\nabla^2\phi$ and $Pe = \bar{a}_2^2 \bar{u}_c / \bar{M}_\phi \gamma$. The first term represents the non-dimensional representation of chemical potential, in which $Cn = \bar{\zeta}/\bar{H}$ denotes Cahn number that controls the interface thickness. On the other side, the second term represents the dimensionless Péclet number.

*3. Governing equations for electric potential and the expression of electric force*

In the present study, we have chosen a leaky dielectric system (LD-LD-LD) system for the numerical simulation. For a leaky dielectric system, the distribution of electric potential is obtained by solving the following governing equation [27]

$$\bar{\nabla} \cdot (\bar{\sigma} \bar{\nabla} \varphi) = 0, \tag{3}$$

where $\bar{\sigma}$ is the electrical conductivity of the system as we mentioned earlier. In phase field model, the $\bar{\sigma}$ for a three phase system [$i, j \in (1,3)$] can be expressed in terms of phase field parameter depicted as $\bar{\sigma} = \frac{(1+\phi)}{2}\sigma_i + \frac{(1-\phi)}{2}\sigma_j$. In non-dimensional form, the electric potential equation is written as

$$\nabla \cdot (\sigma \nabla \varphi) = 0, \tag{4}$$

where $\sigma$ in non-dimensional form is written in the following format

$$\sigma = \frac{(1+\phi)}{2} R_{ij} + \frac{(1-\phi)}{2}. \tag{5}$$

The governing equation dealing with the distribution of electric potential satisfies the following boundary condition at the $B_T$ and $B_W$

$$\left. \begin{array}{l} \text{at the upper wall } B_T, \ \varphi_t = \bar{H}\bar{E}, \\ \text{at the bottom wall } B_W, \ \varphi_w = 0, \end{array} \right\} \tag{6}$$

*4. Coupling between phase field and elctrohydrodynamics*

In the realm of phase field method, the pressure and velocity field is obtained via the solution of the continuity equation and Cahn-Hilliard equation. The Cahn-Hilliard equation couples the phase field model with the electrohydrodynamic. The continuity and Cahn-Hilliard equation are represented in the following format

$$\nabla \cdot \mathbf{u} = 0, \; Re\left(\frac{\partial \mathbf{u}}{\partial t} + \nabla \cdot (\mathbf{u}\mathbf{u})\right) = -\nabla p + \nabla \cdot \left[\mu\{\nabla \mathbf{u} + (\nabla \mathbf{u})^T\}\right] + \frac{1}{Ca}G\nabla \phi + \frac{Ca_E}{Ca}\mathbf{F}^E. \qquad (7)$$

In the above mentioned equation (7), $\mathbf{F}^E$ denotes the volumetric electrical force that causes the deformation of the interface and expressed as $\mathbf{F}^E = \nabla \cdot (\varepsilon \nabla \varphi)\nabla \varphi - |\nabla \varphi|^2 \nabla \varepsilon / 2$ [27]. The expression of $\mathbf{F}^E$ consists of two terms. The first term of the expression denotes the electrophoretic force and the second term denotes the dielectrophoretic force. In equation (7), the term $G\nabla \phi$ denotes the phase field parameter-depended interfacial tension force. The fluid properties $\rho$, $\mu$ and $\varepsilon$ are expressed in the following form

$$\left.\begin{array}{l}\rho = \dfrac{(1+\phi)}{2}\rho_r + \dfrac{(1-\phi)}{2}, \\[6pt] \mu = \dfrac{(1+\phi)}{2}\lambda_{ij} + \dfrac{(1-\phi)}{2} \\[6pt] \varepsilon = \dfrac{(1+\phi)}{2}S_{ij} + \dfrac{(1-\phi)}{2}\end{array}\right\} \qquad (8)$$

For solving the above mentioned governing equations, suitable boundary conditions have been chosen at the $B_T$ and $B_B$ (with $\mathbf{n}_s$ symbolizing the normal vector at the channel wall) and they are read as

$$\left.\begin{array}{l}\text{(i) No slip}: \mathbf{u} - (\mathbf{u} \cdot \mathbf{n}_s)\mathbf{n}_s = \mathbf{0}, \\ \text{(ii) No penetration}: \mathbf{u} \cdot \mathbf{n}_s = 0, \\ \text{(iii) No flux}: \mathbf{n}_s \cdot \nabla \phi = 0,\end{array}\right\} \qquad (9)$$

The velocity and pressure field are periodic in the horizontal direction and expressed in the following format

$$\left.\begin{array}{l}\text{(i)} \; \mathbf{u}(\mathbf{x}) = \mathbf{u}(\mathbf{x}+L), \\ \text{(ii)} \; p(\mathbf{x}) = p(\mathbf{x}+L), \\ \text{(iii)} \; \phi(\mathbf{x}) = \phi(\mathbf{x}+L).\end{array}\right\} \qquad (10)$$

For numerical simulation using phase field method, we have used finite element based COMSOL Multiphysics software.

## III. RESULTS AND DISCUSSION

In the present analysis, we have performed a details analysis of the motion and pinch-off of a double emulsion traversing in a confined microchannel under combined presence of transverse electric field and back ground pressure driven flow. In the present analysis, we have considered two LD-LD-LD model having $R_{23} < S_{23}$ & $R_{12} > S_{12}$ defined as System A and $R_{23} > S_{23}$ & $R_{12} < S_{12}$ defined as System B. In the following section, we have studied the temporal variation of droplet eccentricity evaluated by eccentricity index $e^* = xi_{cen} - xo_{cen}$, where $xi_{cen}$ and $xo_{cen}$ denote the area averaged center of inner and outer droplet (shown in Fig. 2) for different electrical parameters. Further we have also explored temporal evolution of the minimum distance between the inner and outer interface quantified by $d_{min}$ ( shown in figure 2) that unveils the pinch-off dynamics of outer droplet. Lastly, we have uncovered different patterns of droplet pinch-off in confined channel for a wide range of electrical parameters.

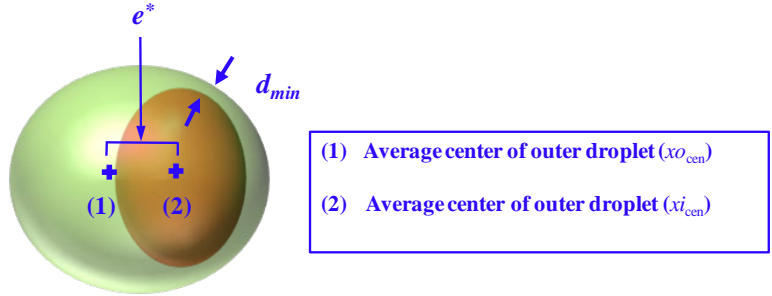

FIG. 2. Simulated parameter used to show the pinch-off of compound droplet: (i) droplet eccentricity ($e^*$), (ii) minimum thickness of the annular zone ($d_{min}$).

## A. Electric field-induced alteration in temporal evolution of droplets eccentricity

Figure 3(a) and 3(b) show the effect of electric capillary number on the temporal variation of droplet eccentricity for system A and system B respectively.

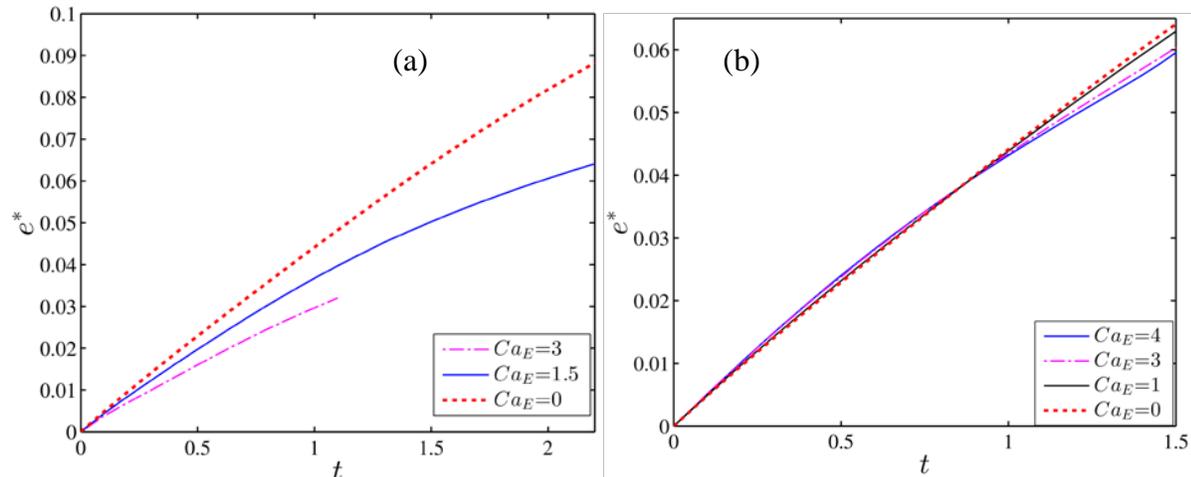

FIG. 3. Effect of $Ca_E$ on the transient evolution of droplet eccentricity for (a) LD-LD -LD system A (b) LD-LD -LD system B. Others parameters are $\lambda=1$, $K=0.67$ and $Re=0.01$.

In solo presence of background pressure driven flow, the inner droplet of a concentric double emulsion moves away from its initial concentric position and creates an eccentric configuration, quantified by eccentricity index $e^*$. Under this condition, the eccentricity of the droplets increases with time monotonically. However, the presence of electric field alters this phenomenon markedly, where the eccentricity-increase rate gradually decreases with the enhancement of electric field strength (or $Ca_E$) for LD-LD-LD system A. We have also observed the same phenomenon for LD-LD-LD system B. However, the effect of electric field is not pronounced unlike the former LD-LD-LD system.

A proper explanation of the observed phenomenon is now provided. When a concentric compound droplet is subjected to background pressure driven flow, the outer droplet experiences more viscous drag compared to the inner droplet due to its larger size and the closeness of wall. Therefore, the inner droplet moves faster as compared to the outer droplet leading to a eccentric configuration with time. It is also worth to mention that, under this condition, the outer droplet deforms in the direction of flow (prolate deformation) and the inner droplet deforms perpendicular to the direction of flow (oblate deformation). For the present LD-LD-LD system A with $S_{23} > R_{23}$ and $S_{12} < R_{12}$, the prolate (or oblate) deformation of the inner (or outer) droplet increases with increase in the electric field strength. Therefore, the annular space at the top and bottom side of the inner droplet get thinner and the fluid in this region gets squeezed, which enhances the pressure in the thin film region as shown in Fig. 4(a). The peak point of the curve in Fig. 4(a) denotes the maximum value of the pressure in the thin annular region. The high pressure in the thin film region reduces the temporal increment in eccentricity by retarding the motion of the inner droplet. On the other hand for LD-LD-LD system B, the outer droplet tries to deform in the direction of electric field and inner droplet deforms perpendicular to it that leads to the formation of thin liquid film near the 'nose' of the outer droplet. Due to the higher pressure in the liquid film (shown in Fig. 4(b)), the motion of the inner droplet is retarded that reduces the temporal increase of droplet eccentricity. One must acknowledge that, for system B, the enhancement of the pressure in the thin annular region is not so pronounced unlike system A. Therefore the effect of electric field on temporal variation of droplet eccentricity is not so significant for system B unlike system A.

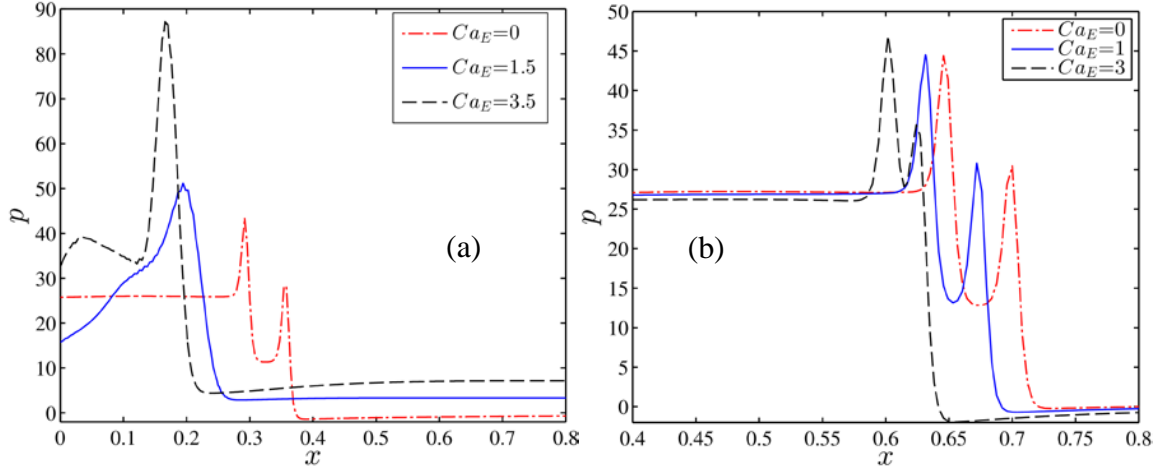

FIG. 4. Variation of pressure in the thin annular region for (a) LD-LD -LD system A (b) LD-LD -LD system B. Others parameters are $\lambda=1$, $K=0.67$, $Ca=0.3$ and $Re=0.01$.

Now, we have discussed about the effect of electrical conductivity of the system ($R_{23}$) on the temporal variation of the droplets eccentricity ($e^*$). From Fig. 5(a), it is well understood that the variation in $R_{23}$ shows a notable effect on the transient variation of droplet eccentricity ($e^*$). In the range of $R_{23} < S_{23}$, increase in $R_{23}$ enhances the temporal increment of $e^*$. However, the value of $R_{23}$ in the range of $R_{23} > S_{23}$ again reduces the temporal increment of $e^*$.

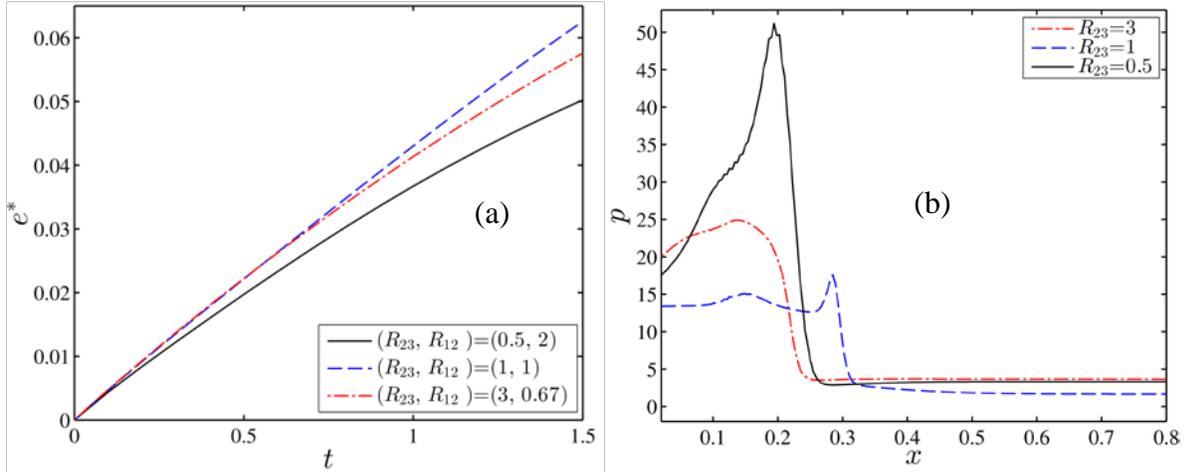

FIG. 5. Effect of conductivity ratio on (a) the transient evolution of droplet eccentricity and (b) pressure distribution in the annular region for LD-LD-LD system having $Ca_E = 1.5$, ($S_{23}$, $S_{12}$)=(2, 0.5). Others parameters are $Ca=0.3$, $\lambda=1$, $K=0.67$ and $Re=0.01$.

.

This is happened due to the fact that, with increase in $R_{23}$ (in the range of $R_{23}<S_{23}$) of the system, the oblate (or prolate) configuration of the outer droplet (or inner droplet) decreases. Therefore, the minimum thickness of the liquid film in the frontal side of the outer droplet increases which reduces the magnitude of developed pressures as depicted in Fig. 5(b). Hence the temporal increases of droplet eccentricity increases. However for higher values of $R_{23}$ (in the range of $R_{23} > S_{23}$), the prolate (or oblate) deformation of outer droplet (or inner droplet) increases that again raises the magnitude of pressure developed at the thin annular region as shown in Fig. 5(b). Therefore, the temporal increment in droplet eccentricity again decreases.

## B. Electric field-induced alteration in pinch-off of droplets

*1. Effect of electric capillary number on the pinch-off liquid droplet*

Figure 6(a) shows the temporal evolution of $d_{min}$ for different values of $Ca_E$ for LD-LD-LD system A. From the figure, it is cleared that, with increase in the electric field strength, the pinch-off of the outer shell occurs at a faster rate. This is happened due to the fact that, with increase in the electric field strength, the oblate (or prolate) deformation of the outer (or inner) droplet increases at a faster rate that reduces the minimum separating distance between the inner and outer interface rapidly. Hence, the rupture of the outer droplet occurs in shorter time. Next, we have find that the temporal variation of $d_{min}$ follows a power law expression as $d_{min} \sim \tau^{\alpha}$, where $\tau$ is expressed as $\tau = t_{rupture} - t$ and $\alpha$ is the exponent of power law. One must acknowledge that the presence of electric field markedly affects the magnitude of $\alpha$. Figure 6(b) shows the thinning behavior of the compound droplet for $Ca_E = 0$ and $Ca_E = 3.5$.

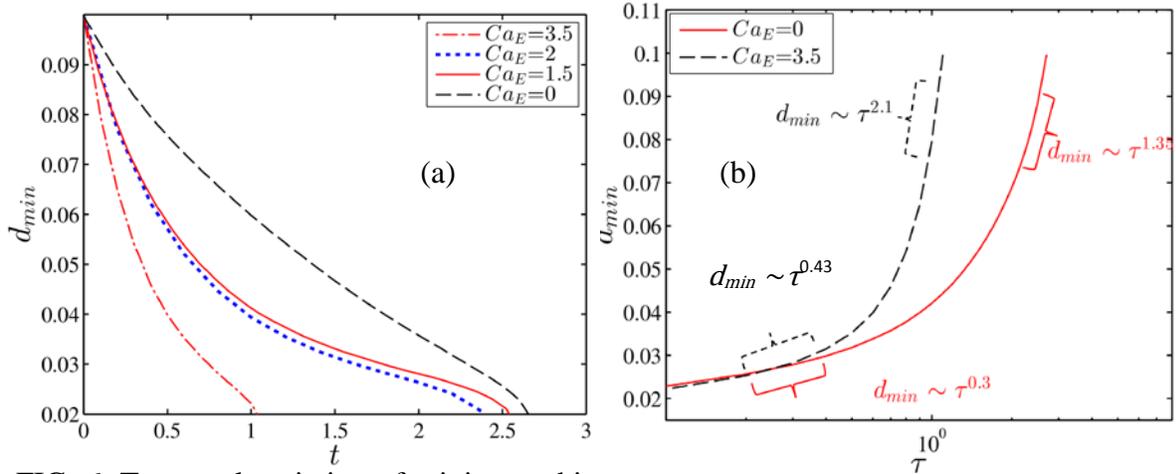

FIG. 6. Temporal variation of minimum thickness between the inner and outer interface of droplet for LD-LD-LD system for different values of $Ca_E$, (b) Relation between minimum thickness and time before rupture for different $Ca_E$. Others parameters are $Ca=0.3,$, $\lambda=1$, $K=0.67$, $(S_{23}, R_{23})=(2, 0.5)$, $(S_{12}, R_{12})=(0.5, 2)$, and $Re=0.01$.

From the figure, it is obtained that both in absence and presence of electric field, the thinning behavior of the system occurs very rapidly at the initial stage of the droplet motion ($\alpha = 1.35$ for $Ca_E = 0$ and $\alpha = 2.1$ for $Ca_E = 3.5$). However beyond a critical minimum thickness of the interface, the thinning rate decreases ($\alpha = 0.43$ for $Ca_E = 3.5$ and $\alpha = 0.3$ for $Ca_E = 0$). The reason is that the squeezed fluid in thin annular region generates a high pressure as shown in Fig. 4(a) that retards the motion of the inner droplet and decreases the values of $\alpha$. The most important thing that needs to be noted that in both regions (initial stage of droplet motion and thinning region), the magnitude of *α* is higher for the case when electric field is present. This is happened due to the higher increasing rate of prolate (or oblate) deformation of the inner (or outer) droplet.

Figure 7(a) also shows a similar phenomenon for LD-LD-LD system B, where the rupture time decreases with increase in the electric capillary number. Furthermore, from Fig. 5(b), it is also obtained that the magnitude of *α* increases with increase in the magnitude of electric field strength both in the initial stage of droplet motion (*α*= 1.35 for *Ca$_E$*=0 and *α*= 1.65 for *Ca$_E$*=4) as well as in the thinning region (*α*= 0.3 for *Ca$_E$*=0 and *α*= 0.48 for *Ca$_E$*=4).

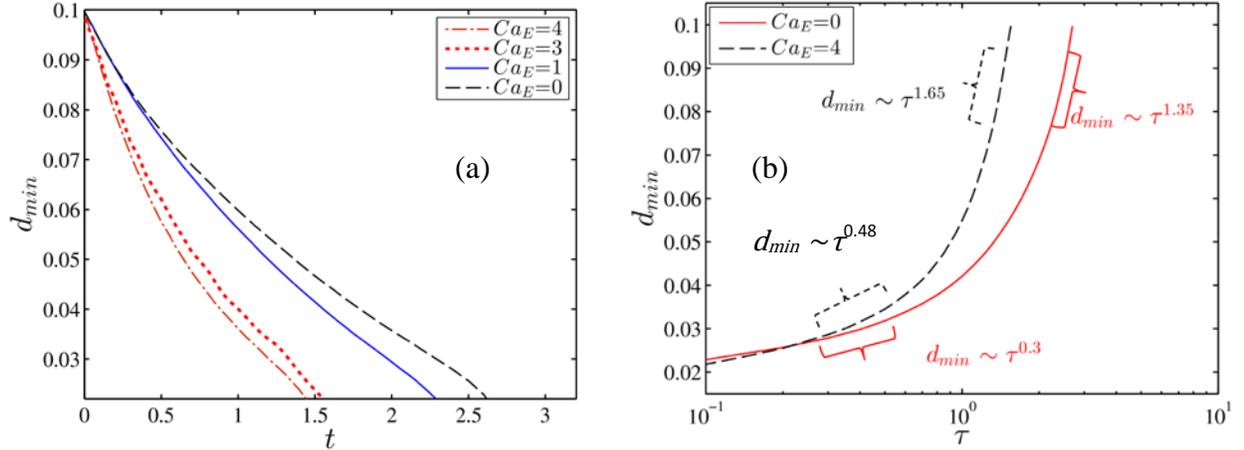

FIG. 7. Temporal variation of minimum thickness between the inner and outer interface of droplet for LD-LD system for different values of *Ca*$_E$, (b) Relation between minimum thickness and time before rupture for different *Ca*$_E$. Others parameters are *Ca*=0.3,, λ=1, *K*=0.67, (*S$_{23}$, R$_{23}$*)=(0.5, 2), (*S$_{12}$, R$_{12}$*)=(2, 0.5) and *Re*=0.01.

Now we have discussed about the effect of electric field on the pattern of droplet pinch off. Under different tested parameter, we have identified two modes of droplet pinch off: (i) hole puncturing the equatorial and (ii) equatorial cap breaking off. In the former mode of break up, a hole is nucleated at the equator of the outer droplet, where as in the later one the outer droplet disintegrates creating a cap liked daughter droplet in the equator. Figure 8 shows the pattern of droplet pinch-off for a LD-LD-LD system A. It is cleared from the figure that, in absence of electric field, the outer shell deforms into a equatorial cap breaking off. In presence of electric field, the break up modes remains unaltered, however the volume of the cap (daughter droplet) increases with increase in the electric field strength. Another important fact is that the location of rupture gradually shifts towards the pole of the outer droplet with the enhancement of electric field strength. This can be attributed to the fact that owing to the prolate and oblate deformation of the inner and outer droplet, a 'shoulder' is created by the inner droplet, where the film thickness is minimum. Further extension of the inner droplet creates a pinch-off the outer shell at the 'shoulder' position. With increase in the electric field strength, the location of the 'shoulder' is shifted towards the poles that also automatically shifts the rupture location of the outer droplet in the same direction and increases the volume of the daughter droplet.

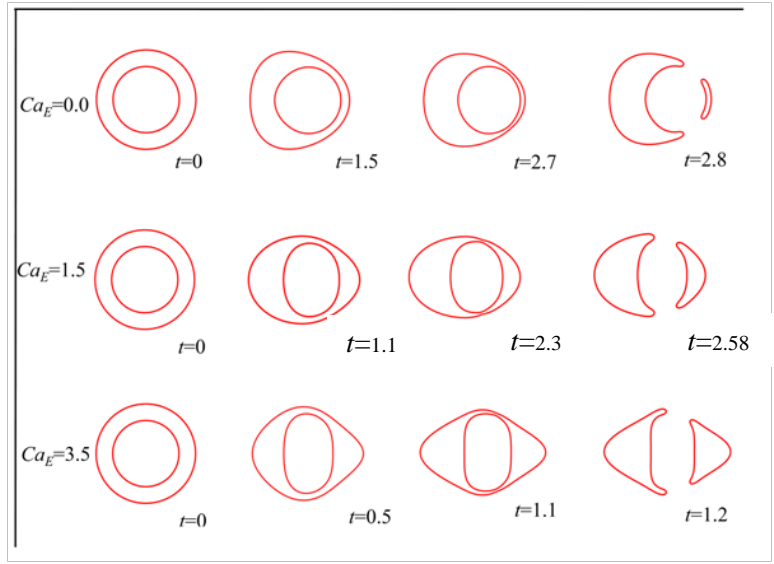

FIG. 8. Effect of electric field on the patterns of droplet rupture Others parameters are $Ca=0.3$,, $\lambda=1$, $K=0.67$, $(S_{23}, R_{23})=(2, 0.5)$, $(S_{12}, R_{12})=(0.5, 2)$, and $Re=0.01$.

Unlike the LD-LD-LD system A, the scenario gets completely changed for a LD-LD-LD system B where the equatorial cap breaking off mode of the outer droplet is gradually converted to hole puncturing the equatorial mode with the increase in electric field strength as shown in Fig. 9.

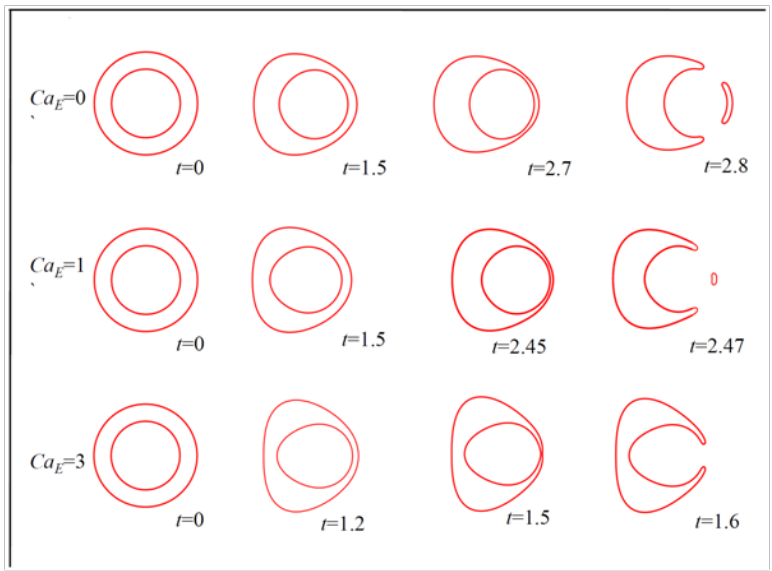

FIG. 9. Effect of electric field on the patterns of droplet rupture Others parameters are $Ca=0.3$,, $\lambda=1$, $K=0.67$, $(S_{23}, R_{23})=(0.5, 2)$, $(S_{12}, R_{12})=(2, 0.5)$ and $Re=0.01$.

Another important point needs to be noted that for the present LD-LD-LD system, with increase in the electric field strength, the prolate (or oblate ) deformation of the outer ( or inner) droplet increases that shifts the location of the rupture of the outer droplet toward its poles. Due to the shifting of the rupture points towards the pole, the cap volume (daughter droplet) decreases and finally the two rupture points converge to a single point creating a hole at the equator. Therefore, the hole puncturing break up mode is observed.

*2. Effect of electrical conductivity on the pinch-off liquid droplet*

In this section, we have discussed about the effect of electrical conductibility of the system on the pinch-off of the composite system. Figure 10(a) shows that the rupture time of the outer droplet increases with increase in electrical conductivity of the system. Figure 10(b) shows that, in the early stage of droplet motion, the thinning rate is much higher for lower values of *R* ($\alpha$ =2.98 for *R*=0.5 and $\alpha$ =1.4 for *R*=3), where as in the thinning region, the thinning rate is higher for higher values of *R* and finally the rupture takes place in shorter time for lower values of *R*. At the early stage of droplet motion, the prolate deformation of the inner droplet and the oblate deformation of the outer droplet increases rapidly for lower values of *R* that reduces the minimum distance between the interfaces very rapidly. However in the thinning region, the pressure developed is higher for lower values of *R* as shown in Fig. 5(b) that leads to the reduction of thinning rate. It is also important to note from Fig. 10(c) that the outer droplet disintegrates with equatorial cap breaking off for lower values of conductivity ratio. With slightly increase in the values of *R*, the volume of the cap (daughter droplet) decreases, but the break up modes remains unaltered. Further enhancement of the values of *R* again increases the volume of the daughter droplet. For lower values of *R*, the oblate deformation of the outer droplet and the prolate deformation of the inner droplet is more. Due to that reason, the contact between the outer and inner droplet occurs near the poles and the rupture points shifts towards the poles of the outer droplet. Hence the volume of the daughter droplet increases. With increase in the value *R*, the oblate deformation of the outer droplet and the prolate deformation of the inner droplet decreases that shifts the rupture location towards the equator and generates a daughter droplet having smaller volume. With further increases in the values of *R*, the outer droplet tries to deform in the direction of electric field where as the inner droplet tries to deform in the direction of flow and again the rupture points is shifted towards the pole that again increases the volume of daughter droplet.

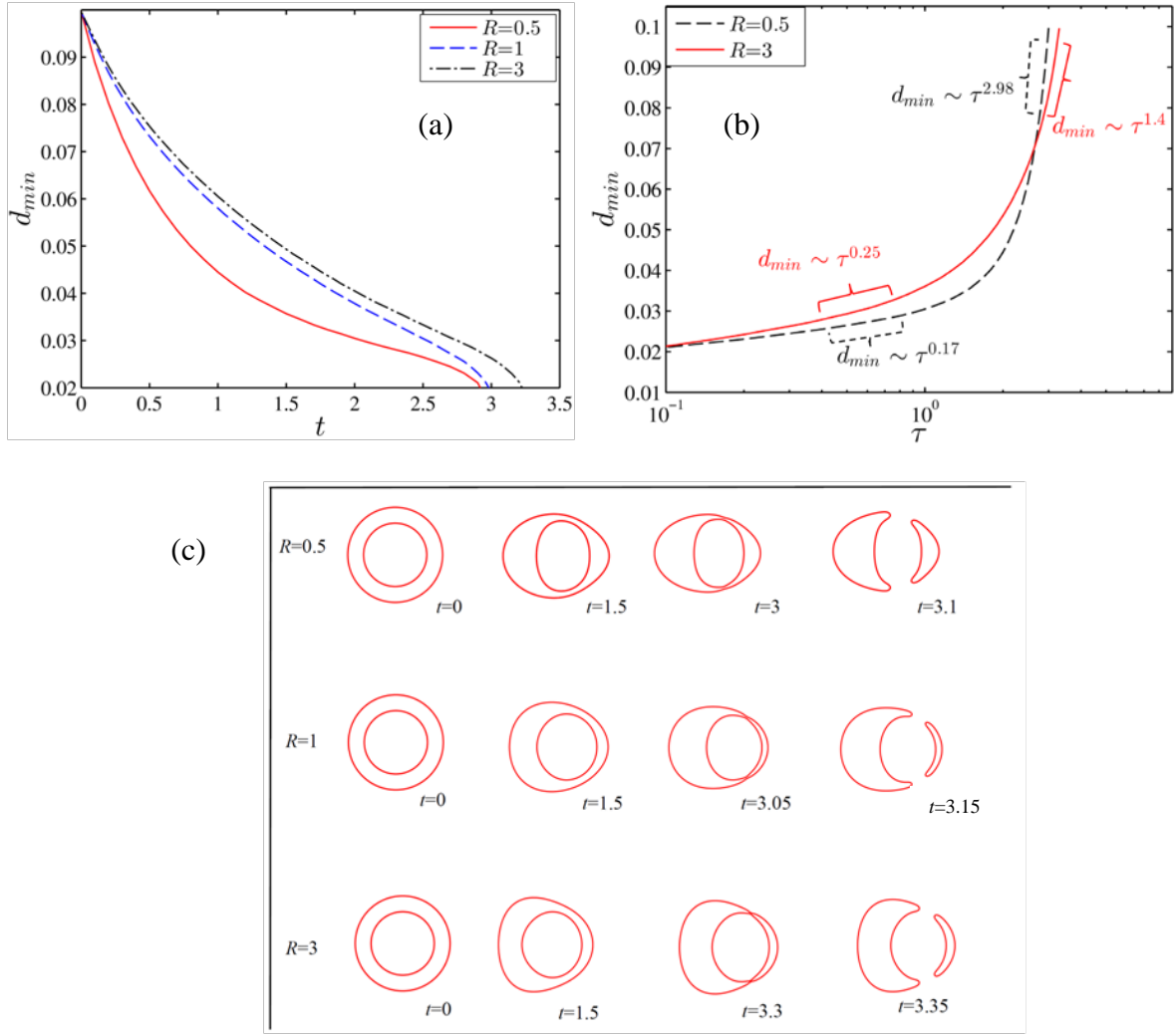

FIG. 10. Temporal variation of minimum thickness between the inner and outer interface of droplet for LD-LD system for different values of $R$, (b) Relation between minimum thickness and time before rupture for different $R$ and (c) effect of electric field on the patterns of droplet rupture. Others parameters are $Ca=0.3$, $Ca_E = 1.5$, $\lambda=1$, $K=0.67$, $(S_{23}, S_{12})=(2, 0.5)$, $R= R_{23} = 1/R_{12}$ and $Re=0.01$.

*3. Effect of electrical permittivity on the pinch-off liquid droplet*

Figure 11(a) shows the effect of electrical permittivity on the thinning behavior of the outer droplet, where we have plotted the evolution of $d_{\min}$ with time for different values of electrical permittivity.

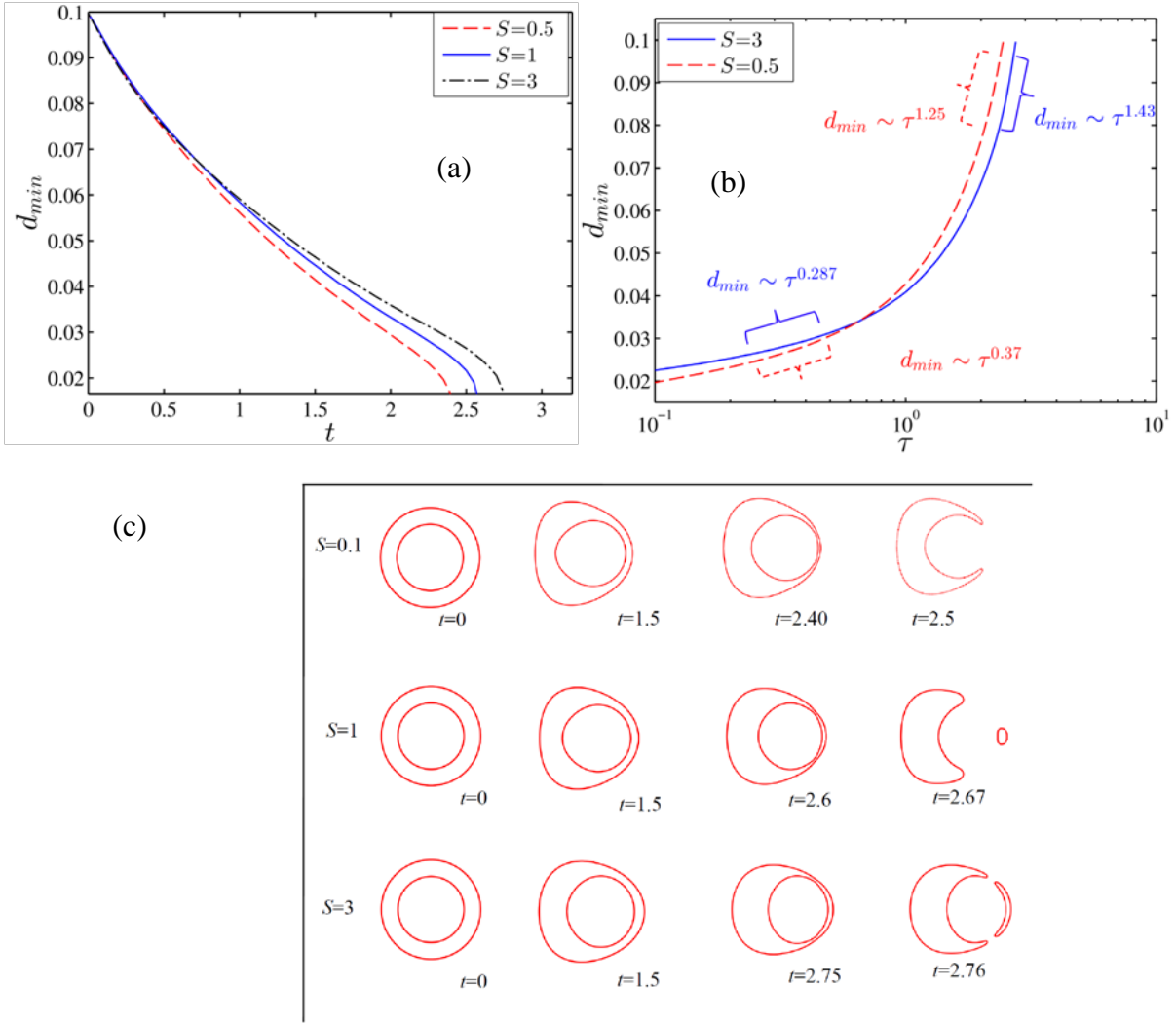

FIG. 11. Temporal variation of minimum thickness between the inner and outer interface of droplet for LD-LD system for different values of $R$, (b) Relation between minimum thickness and time before rupture for different $R$ and (c) effect of electric field on the patterns of droplet rupture. Others parameters are $Ca=0.3$, $Ca_E=1$, $\lambda=1$, $K=0.67$, $(R_{23}, R_{12})=(2, 0.5)$, $S= S_{23} = 1/S_{12}$ and $Re=0.01$.

From the figure, it is clearly understood that the pinch-off time increases with increase in the permittivity ratio of the system. Figure 11(b) shows that the thinning rate increases with increase in the value of permittivity ratio ( $\alpha =1.25$ for $S=0.5$ and $\alpha =1.43$ for $S=3$) at the initial stage of droplet motion. The reason is that, at initial stage of droplet motion, the increase in the oblate (or prolate) deformation of the outer (or inner) droplet with time is more for higher values of permittivity ratio that reduces the minimum distance between the two interfaces very rapidly.

However in highly thinning region, the thinning rate is more for lower values of permittivity ratio due to comparatively lower strength of developed pressure in the annular thin fluid region as shown in Fig. 12 and ultimately the rupture phenomenon takes shorter time for the system with lower values of permittivity ratio.

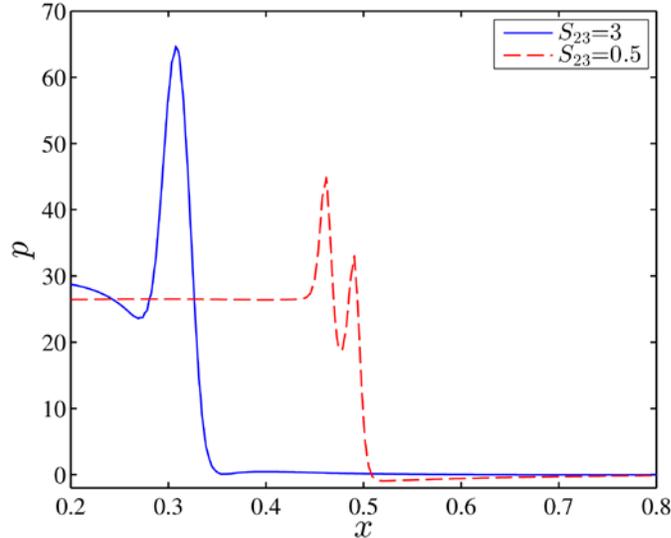

FIG. 12. Effect of permittivity ratio on the pressure distribution in the annular region. Others parameters are $Ca=0.3$, $Ca_E = 1$, $\lambda=1$, $K=0.67$, $(R_{23}, R_{12})=(2, 0.5)$, $S= S_{23} = 1/S_{12}$ and $Re=0.01$.

Another important fact observed in Fig. 11(c) is that for lower values of permittivity ratio, the outer droplet disintegrates with a hole puncturing at the equator and this mode of break up is converted to the equatorial cap breaking off mode for slight increase in the magnitude of permittivity ratio. Further increase in the values of the permittivity ratio enhances the volume of the daughter droplet. At lower values of permittivity ratio, the inner droplet deforms into prolate configuration and the outer droplet deforms into oblate deformation. Therefore the tip of the inner droplet comes into contact with the outer droplet at equator and nucleates a hole at the equator. However, with increase in the value of permittivity ratio, the prolate ( or oblate) deformation of the inner (or outer) droplet decreases that shifts the thin annular region towards the poles. Therefore, the rupture location also shifts towards the poles and the outer droplet disintegrates with making a cap at the equator (equatorial cap breaking off mode). Further increase in the values of permittivity ratio shifts the rupture location more towards the pole that increases the volume of the daughter droplet

Next, we have constructed a regime diagram as depicted in Fig. 13 that shows two distinct regimes of droplet pinch-off in confined domain. The regime containing square markers with yellow face color denotes the values of $(R, S)$, at which the droplet undergoes equatorial cap breaking off. On the other hand, the regime containing circular markers with green face color shows the values of (R, S), at which the outer droplet disintegrates into hole puncturing at the equator.

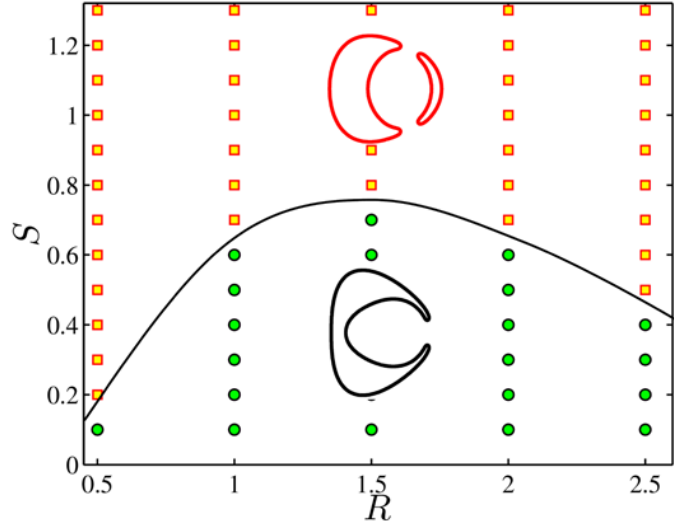

FIG. 13. Regime diagram showing different modes of droplet break up based on the values of (R, S). Others parameters are $Ca=0.3$, $Ca_E = 3$, $\lambda=1$, $K=0.67$, $Re=0.01$, $R=R_{23}=1/R_{12}$ and $S=S_{23}=1/S_{12}$

From Fig. 13, it is also obtained that the critical values of S above which the transition from one break up mode to another break up is taken place first increases with R ( up to R=1.5), then decreases.

**IV. CONCLUSIONS**

The current analysis explored the non-trivial pinch-off dynamics of a compound droplet in a confined medium undergoing pressure driven flow in the presence of a transverse electric field. A finite element based numerical simulation is performed for capturing the essential dynamics of droplet motion and pinch off. Some of the noteworthy results encountered in this analysis are stated below,

(i) For both the LD-LD-LD system having $S_{23} < R_{23}$ & $S_{12} > R_{12}$ and $S_{23} > R_{23}$ & $S_{12} < R_{12}$, with increase in the electric field strength, the temporal increment in eccentricity decreases. However the effect of electric field is pronounced for the latter case.

(ii) For a LD-LD-LD system, initially having $S_{23} > R_{23}$ & $S_{12} < R_{12}$, increase in the $R_{23}$ first increases the temporal increment in the eccentricity of the system and then decreases.

(iii) For the present LD-LD-LD system, the thinning rate increases with increase in the electric field that leads to lower rupture time.

(iii) In absence of electric field, the outer droplet disintegrates from the frontal side creating a cap liked daughter droplet, termed as 'equatorial cap breaking off'. For LD-LD-LD system having $S_{23} > R_{23}$ & $S_{12} < R_{12}$, with increase in the electric field, the breakup mode remains unaltered. However the higher electric field strength shifts the rupture positions towards the pole and increases the volume of the daughter droplet. On the other side, for LD-LD-LD system having $S_{23} < R_{23}$ & $S_{12} > R_{12}$, with increase in the electric field strength the rupture position shifts towards equator and at higher electric field strength, the outer droplet disintegrates by making a hole at the equator termed as 'hole puncturing at the equator'.

(iv) With increase in the values of the conductivity ratio, the rupture time increases. One interesting fact is that, in the initial stage of droplet motion, the increase in the conductivity ratio decreases the thinning rate of the system, where as in the thin region it increases the thinning rate. Furthermore, the rupture point first shifts towards the equator with increase in the values of conductivity ratio. Therefore, the volume of the cap (daughter droplet) decreases. Further extension of the value of conductivity ratio again shifts the pinch-off location towards the pole. Thus produces larger volume of the daughter droplet.

(v) With increase in the values of permittivity ratio, the rupture time of the outer droplet increases. Importantly, higher magnitude of permittivity ratio slow down the thinning behavior at the initial stage droplet motion. However, in the thinning region, it increases the thinning rate. Furthermore at very lower value of permittivity ratio, the outer droplet break off by puncturing a hole at the equator. With slight increase in the permittivity ratio, the rupture points slightly shifts towards the poles and a cap liked daughter droplet is created.

(v) The critical values of the permittivity ratio, above which the transformation from hole puncturing breakup mode to cap breaking off mode takes place, first increases with the values of conductivity ratio up to $R \approx 1.5$. Further extension in the values of $R$ again decreases the values of critical permittivity ratio.

**ACKNOWLEDGMENTS**

S.S. and S.D. are grateful to Dr. Shubhadeep Mandal for insightful discussions on droplet EHD. S.S. and S.D. are grateful to Mr. Ankit Agarwal for insightful discussion on post processing of numerical data.

**APPENDIX A: MODEL VALIDATION STUDY**

For verifying the correctness of the present numerical results, we have performed validation test of our obtained result with the previously published works of Mortazavi and Tryggvason [30] and Halim & Esmaeli [31]. In the first study, we have matched the temporal variation of the lateral position of a deformable droplet in plan poiseuille flow under different conditions with the study of Mortazavi and Tryggvason [30] as shown in Fig.14(a). Next, we have also compared the temporal evolution of deformation parameter of a leaky dielectric droplet suspending in another leaky dielectric medium under a uniform transverse electric field with the result of Halim and Esmaeeli [31] as depicted in Fig. 14(b), Both the figures, Fig. 14(a) and Fig. 14(b) show a very negligible difference between our obtained results and the result obtained by Mortazavi & Tryggvason and Halim & Esmaeeli.

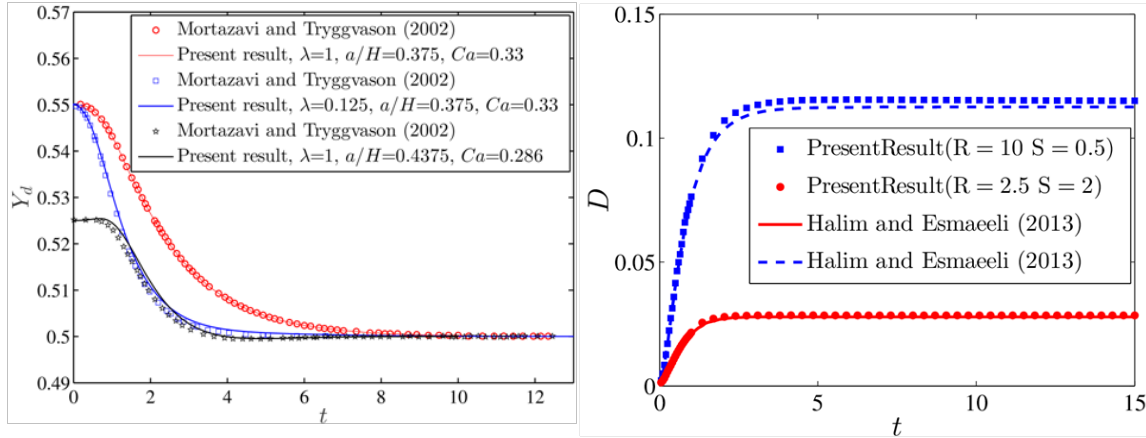

FIG. 14. (a) Variation lateral position of the droplet with time.(b) Variation of deformation parameter with time .Other parameter is *Re* = 0.1

**APPENDIX B: GRID INDEPENDENCE AND CAHN NUMBER INDEPENDENCE STUDY**

For confirming the accuracy of the obtained numerical result and showing the independency of the obtained result on the grid element size, we have carried out the grid independence study. As the Cahn number (*Cn*) and grid size are identical throughout the domain, a *Cn* independence study automatically satisfies a grid independence study. *Cn* = 0.015, 0.01, 0.005 have been taken into consideration for the present study as shown in Fig. 15. From the figure, it can be concluded that there is no significant difference in the droplet trajectory for the chosen different values of *Cn* number. For the present simulation, we have taken *Cn* = 0.01 and all the plots are drawn taking *Cn* = 0.01.

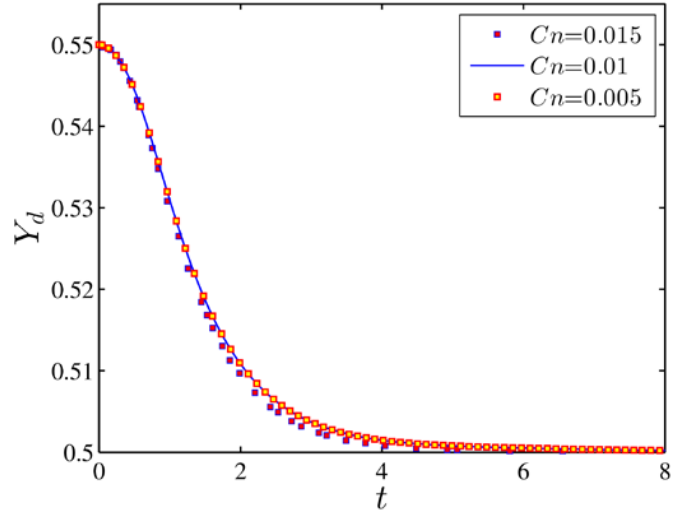

FIG. 15. Grid independence study of temporal variation of the lateral position of the droplet for three different values of *Cn*. Other parameters are *Re*=1, *Ca*=0.33, *a/H*=0.375 and λ=0.125

**References**

[1]   W. a. Macky, Proc. R. Soc. A Math. Phys. Eng. Sci. **133**, 565 (1931).

[2]   C.G Garton and Z. Krasucki, Proc. R. Soc. A Math. Phys. Eng. Sci. **280**, 211 (1964).

[3]   S. L. Anna, Annu. Rev. Fluid Mech. **48**, 285 (2016).

[4]   S.-Y. Teh, R. Lin, L.-H. Hung, and A. P. Lee, Lab Chip **8**, 198 (2008).

[5]   Y. Zhu and Q. Fang, Anal. Chim. Acta **787**, 24 (2013).

[6]   S. Mhatre, V. Vivacqua, M. Ghadiri, A. M. Abdullah, A. Hassanpour, B. Hewakandamby, B. Azzopardi, and B. Kermani, Chem. Eng. Res. Des. **96**, 177 (2015).

[7]   H. C. Shum, Y. Zhao, S.-H. Kim, and D. A. Weitz, Angew. Chemie **123**, 1686 (2011).

[8]   D. Lee and D. A. Weitz, Adv. Mater. **20**, 3498 (2008).

[9]   M. J. Blanco-Prieto, E. Fattal, A. Gulik, J. C. Dedieu, B. P. Roques, and P. Couvreur, J. Control. Release **43**, 81 (1997).

[10]  H. Okochi and M. Nakano, Adv. Drug Deliv. Rev. **45**, 5 (2000).

[11]  A. S. Utada, Science (80-. ). **308**, 537 (2005).

[12]  Y. Zhang, H. F. Chan, and K. W. Leong, Adv. Drug Deliv. Rev. **65**, 104 (2013).

[13]  H.-C. Kan, H. S. Udaykumar, W. Shyy, and R. Tran-Son-Tay, Phys. Fluids **10**, 760 (1998).

[14]   S. Tasoglu, G. Kaynak, A. J. Szeri, U. Demirci, and M. Muradoglu, Phys. Fluids **22**, 82103 (2010).

[15]   J. F. Edd, D. Di Carlo, K. J. Humphry, S. Köster, D. Irimia, D. A. Weitz, and M. Toner, Lab Chip **8**, 1262 (2008).

[16]   M. P. Borthakur, G. Biswas, and D. Bandyopadhyay, Phys. Rev. E **97**, 43112 (2018).

[17]   H. N. Gouz and S. S. Sadhal, Q. J. Mech. Appl. Math. **42**, 65 (1989).

[18]   T. Tsukada, J. Mayama, M. Sato, and M. Hozawa, J. Chem. Eng. JAPAN **30**, 215 (1997).

[19]   J.-W. Ha and S.-M. Yang, Phys. Fluids **11**, 1029 (1999).

[20]   A. Behjatian and A. Esmaeeli, Acta Mech. **226**, 2581 (2015).

[21]   A. Behjatian and A. Esmaeeli, Phys. Rev. E **88**, 33012 (2013).

[22]   P. Soni, V. A. Juvekar, and V. M. Naik, J. Electrostat. **71**, 471 (2013).

[23]   P. Soni, D. Dixit, and V. A. Juvekar, Phys. Fluids **29**, 112108 (2017).

[24]   P. Soni, R. M. Thaokar, and V. A. Juvekar, Phys. Fluids **30**, 32102 (2018).

[25]   V. E. Badalassi, H. D. Ceniceros, and S. Banerjee, J. Comput. Phys. **190**, 371 (2003).

[26]   D. Jacqmin, J. Comput. Phys. **155**, 96 (1999).

[27]   S. Santra, S. Mandal, and S. Chakraborty, Phys. Fluids **30**, 62003 (2018).

[28]   S. Mandal, U. Ghosh, and A. Bandopadhyay, J. Fluid Mech. **776**, 390 (2017).

[29]   P. K. Mondal, U. Ghosh, A. Bandopadhyay, D. DasGupta, and S. Chakraborty, Phys. Rev. E **88**, 23022 (2013).

[30]   S. Mortazavi and G. Tryggvason, J. Fluid Mech. **411**, S0022112099008204 (2000).

[31]   M. A. Halim and A. Esmaeeli, Fluid Dyn. Mater. Process. **9**, 435 (2013).